# Magnetic plasmonic Metamaterials in actively pumped Host Medium and Plasmonic Nanolaser

Andrey K. Sarychev [a], Gennady Tartakovsky [b]

[a] Ethertronics, Inc, San Diego, 9605 Scranton Rd., CA 92121, [b] Simultek, San Diego, CA 92130

## ABSTRACT

We consider plasmonic nanoantennas immersed in active host medium. Specifically shaped metal nanoantennas can exhibit strong magnetic properties in the optical spectral range due to the excitation of Magnetic Plasmon Resonance (**MPR**). A case when a metamaterial comprising such nanoantennas can demonstrate both "left-handiness" and negative permeability in the optical range is considered. We show that high losses predicted for optical "left-handed" materials can be compensated in the gain medium. Gains required to achieve local generation in such magnetic active metamaterials are calculated for real metals. We propose plasmonic nanolaser, where the metal nanoantenna operates like a resonator. The size of the proposed plasmonic laser is much smaller than the wavelength. Therefore, it can serve as a very compact source of EM radiation.

## 1. Introduction

Extending the range of electromagnetic properties of naturally occurring materials motivates the development of artificial metamaterials. For example, it has been demonstrated recently that metamaterials may exhibit such exotic properties as artificial magnetism [1], negative dielectric permittivity (see, for example [1,2]), negative magnetic permeability [3], and even both [4-6]. The double-negative case of Re ε < 0 and Re μ < 0 is often referred as a left-handed material (LHM). Situations when a negative refractive index can be realized in practice are particularly interesting because of the possibility of a "perfect" lens with subwavelength spatial resolution [6]. In addition to the super resolution not limited by classical diffraction, many unusual and sometimes counter-intuitive properties of negative refraction index materials (NIM) make them very promising for applications in resonators, waveguides, and other microwave and optical elements [7–10]. Negative refraction and subwavelength imaging has been demonstrated in the microwave regime [5,8,9,11]. For microwave NIMs, artificial magnetic elements providing Re μ < 0 are the resonators of the split ring type. In the microwave range of the spectrum metals can be considered perfect conductors because the skin depth is much smaller than the metallic feature size. The strong magnetic response is achieved by operating in the vicinity of the LC resonance of the split ring [1,3,12]. The same technique of obtaining Re μ < 0 using split rings was recently extended to mid-IR [12] by scaling down the dimensions of the split rings.

For the microwave and mid-IR wavelengths, metals can be treated as perfect conductors. Therefore, the frequencies of the LC resonances are determined entirely by the split ring geometry and size, not by the electromagnetic properties of the metal. In accordance with this statement, the ring response is resonantly enhanced at some particular ratio of the radiation wavelength and the structure size. Thus, we refer to the LC resonances of perfectly conducting metallic structures as geometric LC (GLC) resonances. The situation drastically changes in the optical (near-IR) part of the spectrum, where thin subwavelength metal components behave very differently when their sizes become less than the skin depth. For example, the electrical surface plasmon resonance (SPR) occurs in the optical and near-IR parts of the spectrum due to collective electron oscillations in metal structures. Many important plasmon-enhanced optical phenomena and applications of metal nanocomposites are based on the electrical SPR (see, for example [13]).

Plasmonic nature of the electromagnetic response in metals for optical–mid-IR frequencies is the main reason why the original methodology of GLC resonances in the microwave– mid-IR spectral range is not extendable to higher frequencies. For the optical range, NIMs with a negative refractive index were first demonstrated in [14] where the authors observed the real part of the refractive index $n_1 = -0.3$ at the telecommunication wavelength of 1.5 μm. In [14] paper the authors experimentally verified their earlier theoretical prediction of negative refraction in an array of parallel metal nanorods [15]. The first experimental observation of negative *n* in the optical range was followed by another successful experiment [16]. Note that the losses become progressively important with decreasing the wavelength towards the optical range [17]. Moreover, finite losses inside the LHM superlens could dramatically reduce the resolution of such lens [18] and made a dream of a superlens unattainable.

Plasmonic effects must be correctly accounted for the design of metamaterials with optical magnetism. Below we show that specifically arranged and shaped metal nanoparticles can support along with the electrical SPR also a magnetic plasmon resonance (MPR). The MPR frequency can be made independent of the absolute characteristic structure size $a$. In fact, the only defining parameters of the MPR are the metal permittivity $\varepsilon_m$ and the structure's geometry. Such structures act as optical nanoantennas by concentrating large electric and magnetic energies on a nanoscale even at optical frequencies. The magnetic response is characterized by the magnetic polarizability $\alpha_M$ with the resonant behavior similar to the electric SPR polarizability $\alpha_E$: real part of $\alpha_M$ changes the sign near the resonance and becomes negative, as required for negative index meta-materials.

We show that MPR must replace or strongly modify GLC resonances in the optical/mid-IR range if a strong magnetic response is desired. The idea of MPR for inducing optical magnetism is relatively new [19]. Horseshoe-shaped structures, first suggested in [20] are described below. These structures support strong magnetic moments at frequencies higher than microwave–mid-IR range for which traditional split ring resonators (see [3,12] for details) were proposed and demonstrated. Conceptually, the horseshoe-shaped structures described here are distinct from the previously studied low frequency structures, which relied on the GLC resonance for producing a strong magnetic response. Horseshoe nanoantennas have distinctively different magnetic response from split ring antennas due to their "elongated" shape and concentration of the EM field inside the gap between the "arms" of the horseshoe.

Plasmonic properties of the metal are very important when the sizes are small and the operational frequencies are high. In the next section, we will outline the derivation of the magnetic permeability of the metamaterial comprising such horseshoe shaped metallic nanoparticles. We are interested in the horseshoe resonators filled with an active medium.

## 2. Electrodynamics of Horseshoe shaped Nanoparticles

We consider the interaction of the metallic horseshoe shaped nanoantenna with a two-level amplifying system (TLS), which can be represented by a quantum dot in the semiconductor host or another high gain medium. In this paper, we will try to derive the conditions under which such a medium can demonstrate very low loss or even a gain leading to lasing in the LHM. Lasing in the random medium without cavity was predicted by V.S. Letokhov [21] and first demonstrated experimentally by N. Lawandy *et al.* [22]. Recently strong lasing was demonstrated in the dye solution containing 55 nm silver nanoparticles [23]. Very recently, increased gain up to $10^5$ cm$^{-1}$ was demonstrated in the medium comprising quantum dots in photonic crystal [24, 25]. Propagation of the surface plasmon-polariton in the interface of metal and gain medium was considered in many papers [26-31] Recently the effect of small increase in plasmon amplitude due to interaction with active medium was observed in the experiment [32]. The enhancement of the fluorescence in the dye film deposited on a corrugated silver surface was investigated in [33]. Most recently, when this paper was submitted for publication a new advance in improved gain of SP in active medium was achieved. A six-fold gain demonstrated by Noginov *at al.* in [34]. The possibility of nonradiative transfer of energy from the active medium to the static plasmon modes of a metal nanoparticle was discussed by Bergman and Stockman [35]. Increasing the resolution by removal of absorption in a near-field lens via optical gain was discussed in [36]. Red shift of the excitonic peak and enhanced local field in quantum dots was observed in [37]. The review of recent works on LHM in gain medium is presented in [38]. Below we will consider optical properties of magnetic nanoantennas in active medium with high gain.

For the sake of simplicity, we assume that TLS are placed inside the horseshoe resonator of a type shown in Fig. 1. An external pump provides the population inversion in the TLS. The pump mechanism may be optical or electrical, when carriers are injected from the bands of a semiconductor material surrounding the embedded quantum dot. In the equations, we will characterize only the pump rate and initial population inversion of the TLS interacting with MPR field. First, we will present the calculations of the electromagnetic fields in a system comprising metallic nanoantenas with resonant frequency lying in the visible on NIR band.

We consider a metal nanoantenna that has a shape of horseshoe as shown in Fig.1. For the sake of simplicity of further calculations, we will consider the two-dimensional geometry when the transverse size of the horseshoe is much larger than other dimensions. The external magnetic field excites the electric current in the arms of the horseshoe as shown in Fig. 1. Magnetic moment associated with the circular current that flows in the antenna is responsible for the magnetic response of the nanoantenna. The external magnetic field $H = \{H_0 \exp(-i\omega t), 0, 0\}$ is applied in the plane of the horseshoe (we assume that the electric field and the current circulate in $\{y, z\}$ plane). The circular current $I(z)$ excited by the time-varying magnetic field flows in opposite directions in each arm of the horseshoe as shown in the figure. The displacement currents flowing between horseshoe's arms conclude the close circuit. We neglect the edge effects and assume that the currents and fields are independent of coordinate $x$. To find the currents and the fields we recapitulate the approach of Ref. 19.

Below we will consider nano-horseshoes with size much smaller than the wavelength λ of the external electromagnetic field. Hence, we can introduce a potential difference

$$U(z) = dE_y = 4\pi d \left( Q(z) - P(z) \right) \tag{1}$$

between the arms, where $Q(z)$ is the local electric charge per unite area and $P(z)$ is the polarization of the medium inside the horseshoe. The electric current $I(z)$ generates magnetic field $H(z) = 4\pi I(z)/c$ inside the horseshoe. Therefore, both electric and magnetic fields are present in the nanoantenna, which makes the quasistatic approximation often used in treatment of plasmonic problems inadequate. Below we show the proper approach to the treatment of nanoantennas with magnetic properties. To find the current $I(z)$, we integrate the equation representing the Faraday's law:

$$\operatorname{curl} \mathbf{E} = -\frac{\partial}{c\partial t} \left( \mathbf{H} + \mathbf{H}_0 \right), \tag{2}$$

over the contour $\{a, b, c, d\}$ in Fig. 1, where $\mathbf{H}_0$ is the amplitude of external magnetic field of the electromagnetic wave with wave number $k = \omega/c = 2\pi/\lambda$. It is assumed that the horseshoe length $a$ is much larger than the distance $d$ between the arms.

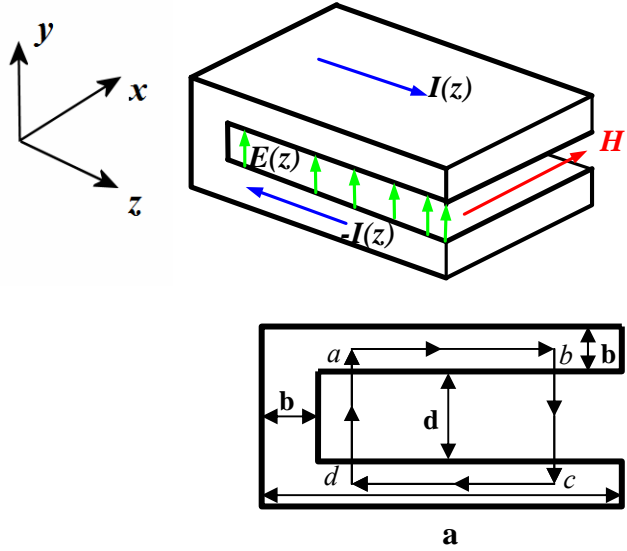

**Figure 1.** Horseshoe nanoantenna geometry; parameters used in modeling: **a** = 300 nm, **d** = 70 nm, **b** = 34 nm.

We also assume that $kd \ll 1$. Integration of (2) results in the following differential equation:

$$(2IZ + \partial U/\partial z)\Delta z = -\frac{d}{c}\left( 4\pi d \, \dot{I}(z)/c + \dot{H}_0 \right) \Delta z, \tag{3}$$

where $\Delta z$ is the distance between points $a$ and $b$ on the integration path in Fig.1, $\dot{I}$ and $\dot{H}_0$ are time derivates, $Z = 1/(\sigma b) = 4i\pi/(\varepsilon_m \omega b)$ is the surface impedance, and $\varepsilon_m = i 4\pi \sigma_m / \omega$ is the metal complex permittivity. By taking time derivative of both sides of Eq. (3) and taking into the account the charge conservation law, $\partial I / \partial z = -\dot{Q}$ we derive the differential equation for the current

$$4\pi\left(\frac{\partial^2 I(z,t)}{\partial^2 z} + \frac{\partial \dot{P}(z,t)}{\partial z}\right) - \frac{2Z}{d}\dot{I}(z,t) = \frac{1}{c}\left(\frac{4\pi}{c}\ddot{I}(z,t) + \ddot{H}_0(t)\right). \tag{4}$$

To find the current and fields in the horseshoe we should add to this equation the constitutive law for the polarization $P$. We assume that the polarization $P$ can be presented as the sum $P = P_1 + P_2$, where $P_1 = \chi_1 E_y$ is a regular, frequency independent polarization, and $P_2$ is the "anomalous" polarization due to the resonant response of TLS system. Typically $P_1 \gg P_2$, this allows us to rewrite Eq. (4) in the following form

$$\frac{\partial^2 I(z,t)}{\partial^2 z} + \frac{\partial \dot{P}_2(z,t)}{\partial z} - \frac{2Z\varepsilon_d}{4\pi d}\dot{I}(z,t) = \frac{\varepsilon_d}{4\pi c}\left(\frac{4\pi}{c}\ddot{I}(z,t) + \ddot{H}_0(t)\right), \tag{5}$$

where $\varepsilon_d = 1 + 4\pi\chi_1$ is the "regular" part of the dielectric permittivity.

We consider first the simplest case when TLS polarizability can be presented as $P_2 = \chi_2 E_y$. Suppose that the external field $H_0$ is harmonic, i.e. $H = H_0 \exp(-i\omega t)$. Then Eq. (4) takes the following form

$$\frac{d^2 I(z)}{d^2 z} = -g^2 I(z) - \frac{\varepsilon_d \omega k}{4\pi} H_0, \tag{6}$$

where $0 < z < a$, $dI(0)/dz = I(a) = 0$, and the parameter $g$ is given as:

$$g^2 = \varepsilon_d k^2 \left[1 - 2\left(k^2 b d \varepsilon_m\right)^{-1}\right], \tag{7}$$

where the dielectric permittivity includes now the regular and resonance parts $\varepsilon_d = 1 + 4\pi(\chi_1 + \chi_2)$. We will consider the limiting case when $(k^2 bd)|\varepsilon_m| \ll 1$, then $g = \sqrt{-2\varepsilon_d/(\varepsilon_m bd)}$ and depends only on the metal permittivity and thickness of the horseshoe arms, but is frequency (wavelength) independent and does not depend on the absolute length of the arms. In order to have a sharp resonance in Eq. (5), the real part of $g^2$ should be positive and imaginary part should be small. Indeed, at IR/visible frequencies $\varepsilon_m$ is negative (with a smaller imaginary part) for typical low loss metals (e.g. Ag, Au, Al, etc.).

Solution of the Eq. (5) for the current yields:

$$I(z) = \frac{\varepsilon_d \omega k}{4\pi g^2} H_0 \left(\frac{\cos(gz)}{\cos(ga)} - 1\right); \tag{8}$$

Magnetic and electric fields induced by the current in the horseshoe nanoparticle can be presented as:

$$H_x(z) = \frac{\varepsilon_d k^2}{g^2} H_0 \left( \frac{\cos(gz)}{\cos(ga)} - 1 \right), \tag{9}$$

$$E_x = 0, \quad E_y(z) = -i\frac{k}{g} H_0 \frac{\sin(gz)}{\cos(ga)} \quad E_z(y,z) = -ikyH_0 \left( \frac{\cos(gz)}{\cos(ga)} - 1 \right). \tag{10}$$

Equations (8) – (10) describe the electric and magnetic fields generated by plasmons excited in the horseshoe by the external magnetic field $H_0$. We would like to emphasize that neither magnetic, nor electric fields are potential in that type of the nanoantenna. It is taken for granted that the magnetic field is solenoidal field. What is more surprising the electric field is also solenoidal in the horseshoe resonator despite the fact that its size is much smaller than the wavelength. Indeed, it follows from Eq. (10) the electric field $E_y$ depends on the coordinate $z$ and electric field $E_z$ depends on the coordinate $y$. The solenoidal electric field is the essence of MPR.

Since the external magnetic field excites circular electric currents in the horseshoe nanoantennas it is naturally to assume that the metamaterial comprising such nanoparticles will have effective magnetic properties. Let us assume that horseshoe nanoantennas with volume density $p$ are organized in a regular cubic lattice. We will calculate the magnetic permeability $\mu$ of such system using the approach developed by A. K. Sarychev and V. M. Shalaev in [13]. Expression for $\mu$ yields

$$\mu = 1 + \frac{p}{aH_0} \int_0^a H_{in} \, dz; \tag{11}$$

where $H_{in}$ is the magnetic field generated by the horseshoe current. Substituting here $H_{in} = H_x(z)$ from Eq. (6), we obtain

$$\mu = 1 + p \frac{k^2 ab\varepsilon_m}{2} \left( \frac{\tan(ga)}{ga} - 1 \right). \tag{12}$$

Therefore, the permeability of such a system has a Lorentz shaped resonance with the central frequency defined by the following condition (see Fig.2):

$$ga = \sqrt{-\frac{2\varepsilon_d a^2}{bd\varepsilon_m}} = \frac{\pi}{2} \tag{13}$$

We are particularly interested in the case when the dielectric host medium filling the space inside the horseshoe is an active high gain medium. This situation can be modeled to the first approximation, by the assumption that imaginary part of $\varepsilon_d$ is negative [30, 31].

One can see from different examples in Fig. 2 that increased gain leads to the narrowing of the absorption line of the horseshoe and eventually the horseshoes start to behave as a set of nanolasers. At the threshold, the permeability becomes singular. In a real system, the permeability will saturate. The process of saturation is disused below in the framework of Maxwell-Bloch equations.

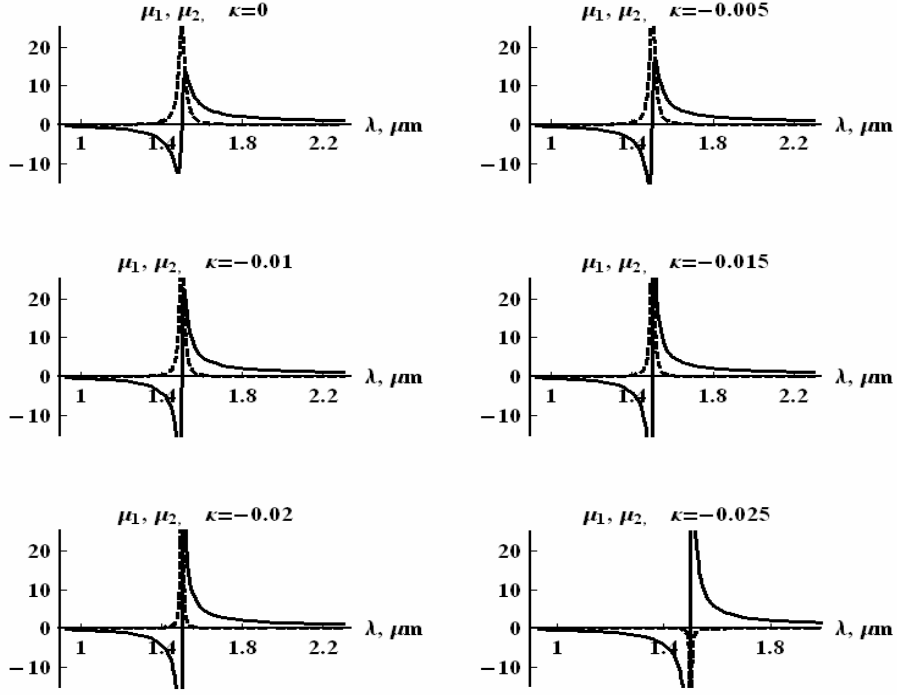

**Figure 2**. Effective permeability of silver horseshoe comprising metamaterial; effective permittivity of actively pumped host medium $\varepsilon_d = 4(1 + i\kappa)$, where the loss factor $\kappa < 0$; parameters used in modeling are: a = 300 nm, d = 70 nm, b = 34 nm; Volume density of horseshoes p=0.3; the real part of permittivity is shown by the dashed lines and imaginary part by solid lines.

## 3. L – C – R model of horseshoe resonator

The main features of horseshoe dynamics can be understood in terms of a simple equivalent model. We still consider the metal horseshoe nanoantenna, which is excited by the magnetic component $H_0$ of the impingent electromagnetic field, as it is shown in Fig. 1. The electric current $I$ flowing in the metal arms of the antenna is shorted by the displacement currents (vertical arrows in Fig. 1). The metal part of the nanoantenna can be presented as an inductance. The gap between two arms is modeled as capacitance. Then the horseshoe antenna can be presented as *L-C-R* circuit, shown in Fig. 3. The inductance $L_a$ stands for the metal since metal's permittivity is typically negative in the optics and IR range and it is proportional to $\omega^{-2}$. The resistance $R_1$ presents the losses in the metal. It always has a positive value.

The resistance $R_2$ stands for the losses in the dielectric, which fills the space between the two arms of the nanoantenna. For the ordinary dielectric material $R_1 > R_2 > 0$. The EMF "generator" $V = V_0 \cos(\omega t)$ in Fig. 3 presents electromotive force induced by the external magnetic field $H_0$.

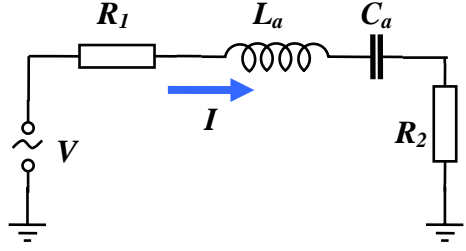

**Figure 3** Equivalent LCR circuit of the horseshoe antenna

For the equivalent circuit in Fig. 3 we obtain the following Kirchhoff's equation, which we write in terms of the electric charge $q$.

$$\frac{L_a}{c^2}\ddot{q} + U + (R_1 + R_2)\dot{q} = V, \tag{14}$$

where $U$ is the potential drop in the capacitor $C_a$, $c$ is the speed of light. In the considered equivalent circuit approximation the potential $U$, which is given by Eq. (1), is independent of the coordinate "z." It can be written as $U = 4\pi d(Q - P_1 - P_2)$, where $Q = q/S \equiv q/(la)$ is the charge density in the capacitor, $l$ is the horseshoe size in "x" direction, $P_1 = \chi_1 E_y$ and $P_2$ are the regular and resonance (gain) polarization of the medium in the capacitor. We introduce the regular capacitance $C_a = \frac{\varepsilon_d S}{4\pi d}$, $(\varepsilon_d = 1 + 4\pi\chi_1)$; then the potential equals to

$$U = \frac{q}{C_a} - \frac{SNp}{C_a}, \tag{15}$$

where $S = la$, $N$ and $p$ are the density and dipole moment of TLS (e.g. quantum dots) correspondingly. Substituting Eq. (15) in Eq. (14), we obtain the equation

$$\frac{L_a}{c^2}\ddot{q} + \frac{q}{C_a} - \frac{SNp}{C_a} + (R_1 + R_2)\dot{q} = V. \tag{16}$$

It is the equation for the charge (current) oscillation in the horseshoe resonator in the presence of gain medium. Note that the TLS dipole moment $p$ in Eq. (16) is the quantum operator.

We first consider the classic approximation when the polarization $P_2 = Np = \chi_2 E_y$ and susceptibility $\chi_2$ does not depend on the electric field. That is we neglect the depletion of the gain medium. Than Eq. (16) takes usual form of Kirchhoff's equation $(L_a/c^2)\ddot{q} + q/C_a + (R_1 + R_2)\dot{q} = V$, where $\varepsilon_d = 1 + 4\pi\chi_1 + 4\pi\chi_2$ includes both regular and "gain" susceptibilities. It is easy to calculate the current $I \equiv \dot{q}$ in the circuit

$$I(t) = V_a \operatorname{Re}\left[\frac{i\omega\exp(-it\omega)}{\omega^2 + i\gamma\omega - 1} + \exp(-\gamma t/2)\frac{(4i\omega\cos t + (4 + \gamma^2 - 2i\gamma\omega)\sin t)}{\omega^2 + i\gamma\omega - 1}\right], \tag{17}$$

where the frequency and time are measured in terms of the resonance frequency $\omega_r = c/\sqrt{L_a C_a}$, parameter $\gamma = (R_1 + R_2)C_a\omega_r$, and $V_a = C_a V_0$.

To estimate $L_a$, $C_a$, $R_1$, and $R_2$ it is convenient to write the metal permittivity as $\varepsilon_m = -|\varepsilon_m|(1 - i\kappa_m)$ and the permittivity of the medium inside the horseshoe as $\varepsilon_d = |\varepsilon_d|(1 + i\kappa_d)$, where we take into account that $\varepsilon_m$ is almost

negative in optics and infrared $\kappa_m \ll 1$. We also suppose that loss or gain is small in the dielectric, i.e. $|\kappa_d| \ll 1$. Then we obtain

$$L_a = \frac{8\pi a}{k^2 |\varepsilon_m| bl}, \quad C_a = \frac{\varepsilon_d al}{4\pi d}, \quad R_1 = \frac{8\pi a \kappa_m}{\omega |\varepsilon_m| bl}, \quad R_2 = \frac{4\pi d \kappa_d}{\omega a \varepsilon_d l} \qquad (18)$$

where we $l$ is the horseshoe size in "x" direction (see Fig. 1). Note that the equation for the resonance frequency $\omega_r = c/\sqrt{L_a C_a}$, obtained in L-C-R model, coincides up to factor $\sqrt{\pi/4}$ with Eq. (13).

From Eqs. (18) we obtain the relaxation parameter $\gamma = \gamma_1 + \gamma_2$ where

$$\gamma_1 = R_1 C_a \omega = \kappa_m (\omega/\omega_r)^2 \simeq \kappa_m, \quad \gamma_2 = R_2 C_a \omega = \kappa_d, \quad \gamma = \kappa_m + \kappa_d. \qquad (19)$$

Second term in Eq. (17) describes the transient process. We assume that at the initial moment the current $I(0)=0$ and the electric charge $q(0)=0$. This term becomes irrelevant when $\gamma > 0$ and properties of the nanoantenna can be described in terms of the complex impedance

$$Z_a = -i\frac{\omega L_a}{c^2} + \frac{i}{\omega C_a} + R_1 + R_2. \qquad (20)$$

The condition $\text{Im} Z_a = 0$ gives the resonance frequency $\omega_r$ whereas the resistances $R_1$ and $R_2$ determine the linewidth. When the frequency $\omega_r$ of the impingent light is larger than the resonant frequency $\omega_r$, the current $I$ inside the horseshoe induces magnetic field $H_{in}$ which direction is opposite to the direction of the field $H_0$ of the impingent wave. Therefore, the metamaterial, composed from the horseshoe nanoantennas have effective magnetic properties at optical frequencies. For $\omega > \omega_r$ the effective magnetic susceptibility is negative and the permeability could be negative when the concentration of the horseshoe nanoantennas is large enough as it is demonstrated in Fig. 2.

Consider now the antenna where the gap in the horseshoe is filled with an active medium, i.e., the capacitor has negative losses $R_2 < 0$ (Eq. 18). For sufficiently large gain the losses in the metal can be overcompensated and parameter $\gamma < 0$. Then it follows from Eq. (17) the electric current $I(t)$ exponentially increases in the antenna. Note that the instability takes place as soon as $\gamma < 0$ while the first term in Eq. (17) is irrelevant now. In the current literature, most of the authors use the approach that is equivalent to consideration the first term in Eq. (17), i.e., they use a complex impedance (effective permittivity) to describe the medium with gain. We see that obtained thus results have limited meaning since the corresponding solution of the Maxwell equations could be unstable.

The electric current and the field in the antenna cannot increase up to infinity due to the saturation of the active medium. To take into account the depletion of the active medium we write the following phenomenological equation for the electric charge $q(t)$ in the horseshoe

$$\ddot{q}(t) + \left[\gamma_1 - \frac{\gamma_2}{1 + \left(\dot{q}(t)\right)^2}\right]\dot{q}(t) + q(t) = V_0(t) \qquad (21)$$

where $\gamma_1 = R_1 C_a \omega_r$ and $\gamma_2 = R_2 C_a \omega_r$. Equation (21) cannot be solved in the general case, however, some predictions can be made immediately: let assume that at the initial moment there is no electric current or charge in the antenna, which will correspond to the initial conditions $I(0)=0$ and $q(0)=0$. When the external magnetic field is turned on, the current will start to increase exponentially according to Eq. (17). Later on the electric current as well as electric charge will saturate at some level due to the depletion of the active medium. In general, the saturation level will be proportional to the amplitude of the magnetic field in the impingent electromagnetic wave. In our model it is denoted as the amplitude $V_0$ of the electromotive force (EMF) in the equivalent LCR circuit depicted in Fig. 3.

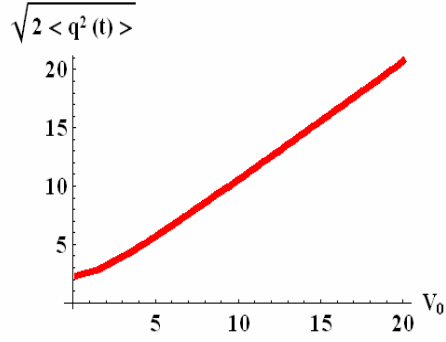

**Figure 4** Averaged solutions of Eq. (21)

The dependence of the charge oscillation amplitude in the horseshoe nanoantenna on EMF $V_0$ is shown in Fig. 4, where the parameters are $\omega = 0.1\omega_r$, $\gamma_1 = 0.1$, $\gamma_2 = -0.5$. We can see that the amplitude increases linearly with sufficiently large $V_0$. It is the region of the forced oscillations when the amplitude is so large that the second term in the square brackets in Eq. (21) is negligible. We also notice (see Fig. 4) the existence of the region of the spontaneous oscillations: the electric current exists even in the absence of external field. Therefore, the horseshoe antenna behaves as a plasmonic nanolaser

## 4. Interaction of Nanoantennas with the active host medium and plasmonic laser

In order to understand the origin of the lasing in active plasmonic medium we will consider a microscopic model following the consideration of a dipole laser presented by Protsenko et al. [39]. We use the quantum-mechanical derivation of the equations of motion for the system shown in Fig. 1, but will neglect quantum correlations and fluctuations in our analysis. The Hamiltonian of the nanoparticle interacting with a TLS is given by the expression:

$$H = H_0 + H_d + V_{int} + \Gamma \qquad (22)$$

where $H_0$ and $H_d$ describe respectively the horseshoe and TLS. The operator $V_{int} = -pE$ gives the interaction between the TLS and the nanoantenna, where $E$ and $p$ are the electric field of the magnetic plasmon excited in the horseshoe and the dipole moment of the TLS respectively. The term $\Gamma$ includes the terms describing dissipation and pump effects.

Electrons couple to the local electric field and oscillate with the frequency $\omega$, which is close to the magnetic plasmon resonance frequency $\omega_r$. We will treat the electric charge $q(t) = q_1(t)\exp(-i\omega t) + q_1^*(t)\exp(i\omega t)$ and the current $I(t) = \dot{q}(t) = I_1(t)\exp(-i\omega t) + I_1^*(t)\exp(i\omega t)$ as classical objects, defined by their slow varying amplitudes $q_1(t)$ and $I_1(t)$. We introduce the operator $b(t) = \eta(t)\exp(i\omega t)$ where the $\eta(t)$ is the operator that corresponds to the transitions between the excited $|e\rangle$ and ground $|g\rangle$ state of the TLS, i.e., $|g\rangle = \eta|e\rangle$ and $\eta|g\rangle = 0$. The operator of the dipole moment in Eq. (22) can be represented as

$$p = \Pi b\exp(-i\omega t) + \Pi^* b^+ \exp(i\omega t) \qquad (23)$$

where $\Pi = \langle g|re|e\rangle$ is the matrix element of the dipole operator between the excited and ground state of TLS. We assume, for simplicity, that all losses are included in term $\Gamma$ in Eq. (22). We assume that the TLS oscillates between the upper and lower level with the frequency $\omega$ close to the frequency $\omega_2$, where $\omega_2$ is the resonance frequency of the TLS.

We express the electric field $E(t)$ in the horseshoe in terms of the electric charge $q(t)$ and the polarization $Np(t)$ of the active medium, where $N$ is the density of TLS. We obtain

$$E(t) = \frac{1}{C_a d}\big(q(t) - SNp(t)\big), \tag{24}$$

where the capacitance $C_a$ of the horseshoe is defined in Eq. (18), $S = al$ is the horseshoe area, and $d$ is the horseshoe gap (see Fig. 1). We will also introduce the population inversion operator

$$D(t) = n_g(t) - n_e(t), \tag{25}$$

where $n_e(t) = b^+ b$ and $n_g(t) = bb^+$ are the operators describing the population of the excited and ground state of the TLS.

Neglecting the fast oscillating terms, $\sim \exp(\pm i2\omega t)$ we can express the Hamiltonian in terms of the following operators:

$$H_d = \hbar\omega_2 n_e, \tag{26a}$$

$$V_{int} = -\frac{1}{C_a d}\big(\Pi^* q_1 b^+ + \Pi q_1^* b\big) + |\Pi|^2 SN, \tag{26b}$$

where the last term in Eq. (26b) is a constant and, therefore, does not influence the dynamic of the system. By using the well-known commutation rules for the operators $b$, $b^+$ and $n_{e,g}$ we derive the equations of motion

$$\dot{b} = -(i\Delta + \Gamma)b + \frac{i}{\hbar C_a d}\Pi^* q_1 D, \tag{27}$$

$$\dot{D} = \frac{2i}{\hbar C_a d}\big(\Pi q_1^* b - \Pi^* q_1 b^+\big) - \frac{D - D_0}{\tau}, \tag{28}$$

where $\Delta = \omega_2 - \omega$. We have included in Eqs. (26) terms involving $1/\tau$ and $\Gamma$ to account for the relaxation and pump processes. $D_0$ is the stationary value of $D$ when $q_1 = 0$. We assume $D_0 < 0$ because the pumping process provides the initial population inversion in the TLS. By neglecting quantum fluctuations and correlation, $D$ and $b$ can be treated as complex variables with $b$, and $b^+$ being replaced by $b$ and $b^*$ respectively.

Thus, we can obtain use Eqs. (16), (27), and (28) as a full set of differential equation that describe the dynamics of the horseshoe nanoantennas in active host medium. We consider now the lasing that is the natural oscillations of the electric charge in the horseshoe resonator in the absence of external field ($V = 0$ in Eq. 16.) We suppose that the resonator is in stationary state so that the amplitudes of oscillation do not change ($\dot{q}_1 = \dot{E}_1 = \dot{b} = \dot{D} = 0$.) Then Eqs. (16) and (27) and (28) can be rewritten the following form

$$(i\delta + \gamma_1)q_2 - ib = 0, \tag{29}$$

$$(i\Delta_1 + \Gamma_1)b - iA_0 D q_2 = 0, \tag{30}$$

$$-\frac{D - D_0}{\tau_1} + 2iA_0\big(q_2^* b - q_2 b^+\big) = 0, \tag{31}$$

where the dimensionless electric charge $q_2 = q_1/(SN\Pi)$, the dimensionless constant

$$A_0 = \frac{4\pi N |\Pi|^2}{\omega_r \hbar \varepsilon_d} > 0, \tag{32}$$

$\Delta_1 = \Delta/\omega_r$, $\Gamma_1 = \Gamma/\omega_r$ and $\delta = (\omega/\omega_r)^2 - 1$, the plasmon resonance frequency $\omega_r$ and the relaxation parameter $\gamma_1$ are given by Eqs. (17) – (19). Equations (29) – (31) define the lasing in the horseshoe plasmonic resonator: they have nonzero solution when the following conditions are fulfilled

$$\frac{\Delta_1}{\Gamma_1} = -\frac{\delta}{\gamma_1}, \tag{33a}$$

$$\left(\frac{\delta}{\gamma_1}\right)^2 + 1 + \frac{A_0 D}{\Gamma_1 \gamma_1} = 0, \tag{33b}$$

The condition (33a) gives the lasing frequency

$$\omega_L = \omega_r + \frac{\gamma_1(\omega_2 - \omega_r)}{\gamma_1 + 2\Gamma_1}, \tag{33c}$$

It always sits between the magnetic plasmon resonance frequency $\omega_r$ and TLS resonance frequency $\omega_2$. In the lasing condition (33b) all terms are positive but the population $D$. Therefore, this condition holds only in the inverted medium $n_e > n_g$ when $D < 0$ (see Eq. 25). The population $D$ cannot be smaller than -1, which corresponds to the case when all TLSs are excited. We obtain the lasing condition for the horseshoe nanolaser

$$\frac{A_0}{\Gamma_1 \gamma_1} = \frac{4\pi N |\Pi|^2}{\hbar \varepsilon_d \Gamma_1 \gamma_1} > 1. \tag{34}$$

As soon as condition (34) is fulfilled the interaction between TLS and the plasmonic nanoantenna leads to coherent oscillations in electric charge and the magnetic moment of the horseshoe, even in the absence of the external electromagnetic field.

To simplify the analysis we consider the system where the plasmon frequency coincides with and TLS frequency, i.e., $\omega_r = \omega_2$ then Eq. (33c) gives the lasing frequency $\omega_L = \omega_r$. We also suppose that the active medium is pumped to such extend that

$$D_0 < -\gamma_1 \Gamma_1 / A_0. \tag{35}$$

Then the amplitude of the charge oscillations $q_2$, TLS polarization $b$, and population $D$ are equal to

$$q_2 = \frac{i e^{i\varphi}}{2 A_0} \sqrt{-\frac{A_0 D_0 + \gamma_1 \Gamma_1}{\gamma_1 \tau_1}}, \tag{36}$$

$$b = -i\gamma_1 q_2, \tag{37}$$

$$D = -\frac{\Gamma_1 \gamma_1}{A_0}, \tag{38}$$

where the amplitudes of the oscillations have arbitrary phase $\varphi$. It is easy to check that Eqs. (36) – (38) indeed give the solution of Eqs. (29) – (31).

The amplitude of the spontaneous electric current in the horseshoe resonator can be estimated as $I_1 = -i\omega_L q_1 = -i\omega_L SN\Pi q_2$. We still assume that $\omega_r = \omega_2 = \omega_L$. Make use of Eqs. (32) and (36) we obtain the electric current

$$I_1 = \frac{\hbar \omega_r^2 e^{i\varphi} S \varepsilon_d}{8\pi \Pi^*} \sqrt{-\frac{A_0 D_0 + \gamma_1 \Gamma_1}{\gamma_1 \tau_1}}, \tag{39}$$

and the magnetic moment of the horseshoe

$$m_1 \simeq \frac{I_1 da}{c} \simeq \frac{\hbar \omega_r^2 e^{i\varphi} aV \varepsilon_d}{8\pi c \Pi^*} \sqrt{-\frac{A_0 D_0 + \gamma_1 \Gamma_1}{\gamma_1 \tau_1}}, \tag{40}$$

where $V = dS$ is the volume of the horseshoe. The magnetic moment exists even in the absence of the external electromagnetic field. Its "direction," i.e., the phase $\varphi$ is undefined in this case.

Let us consider a metamaterial composed of the regular array of the horseshoes nanoantenas filled by active medium. The volume concentration of the horseshoes equals to $p$. We suppose that the lasing condition (35) is fulfilled and interaction between the nanoantenas results in the synchronization, i.e., all the phases $\varphi$ are the same. Then the specific magnetic moment $M_1$ of the metamaterial equals to $M_1 = pm_1/V$. Therefore, the metamaterial has spontaneous high frequency magnetization and operates like an optical ferromagnetic. There is, however an important difference between usual ferromagnetism and proposed optical magnetism. The static ferromagnetism is due to the quantum mechanical, persistent currents and it exists on its own. The optical magnetism exist only if the active medium is pumped: even made from the perfect metal and dielectric the metamaterial radiates when it has spontaneous magnetic moment. The emitted radiation does not need an optical cavity and, in the case of a single horseshoe, could be concentrated in a very small volume $\sim 10^{-2} \, \mu m^3$.

We consider now feasibility of the plasmonic laser and optic magnetism. It is instructive to express the lasing condition (34) in terms of the optical gain. Let us consider the propagation of em wave in the infinite active medium assuming, for simplicity, that the inversion does not change. That is we assume that the population of TLS remains constant $D = D_0$ in the process of em wave propagation. Then the wave propagation can be described in terms of effective dielectric constant $\varepsilon_d + i\varepsilon_d A_0 D/(\Gamma_1 + i\Delta_1)$ (see e.g. Ref. 40). The maximum gain is achieved when the frequency equals to the resonance frequency $\omega_2$ of TLS ($\Delta_1 = 0$) and the active medium is completely inverted ($D = -1$). Then the intensity $|E|^2$ of the wave, propagating along say $x$ direction, exponentially increases $|E(x)|^2 \propto \exp(Gx)$, where the optical gain $G = 2\pi A_0 n_d/(\lambda \Gamma_1)$, and $n_d = \sqrt{\varepsilon_d}$ is nonresonant refractive index of the active medium. Therefore, we can express the lasing condition (34) in terms of the gain

$$\frac{G\lambda}{2\pi n_d \gamma_1} > 1, \qquad (41)$$

recall that $\gamma_1 = \varepsilon_m''/|\varepsilon_m|$ is the metal loss factor. Note that the lasing condition depends only on the gain in active medium and loss in metal. We speculate that Eq. (41) holds for any geometry of subwavelength plasmonic laser. For example, the silver horseshoe would lasing at wavelength 1.5 $\mu m$ if the active medium were had optical gain $G > 3 \cdot 10^3 \, cm^{-1}$. Such gain looks feasible, for example, in quantum dot medium [24,25, see discussion in 30].

## 5. Conclusions

We have shown that horseshoe metallic nanoantennas can be used for building plasmonic metamaterials with both negative permittivity and permeability, which can lead to LHM in NIR and optical band of the spectrum. Contrary to the previously well established believe that most materials do not demonstrate any anomalous magnetic permeability at high optical frequencies, we have shown that specifically shaped nanoparticles can demonstrate negative permeability (and refraction) in optics by exciting MPR. It is well known that plasmonic materials will demonstrate very high absorption in the optical range, which will make them an unlikely candidate for lenses and other practical optical devices. In this paper, we have derived condition under which such nanoantennas filled with highly efficient gain medium can demonstrate low absorption or even gain sufficient for lasing. Proposed host medium should have initial gain greater than $10^3$ cm$^{-1}$. There is still open question regarding the quenching and spontaneous emission of such high gain medium in the near field of plasmonic particles. Set of self-consistent equations derived in this paper allows calculating those effects as well. These results will be presented elsewhere. We would like to express our appreciation to M.A. Noginov for critical reading of this manuscript and bringing to our attention preprint of the paper [34], in which a six-fold enhancement in surface plasmonic resonance in the mixture silver aggregates and pumped R6G dye was observed.